\documentstyle[aps]{revtex}

\begin{document}
\draft
\title{
Tidal Stablization of Neutron Stars and White Dwarfs}

\author{Dong Lai}
\address{Theoretical Astrophysics, 130-33, 
California Institute of Technology\\
Pasadena, CA 91125\\
{\rm E-mail: dong@tapir.caltech.edu}}

\date{\today} 
\date{Accepted for publication in {\it Physical Review Letters}, 1996} 

\maketitle 
\begin{abstract}

What happens to a neutron star or white dwarf near its maximum mass
limit when it is brought into a close binary orbit with a companion?
Such situation may occur in the progenitors of Type Ia supernovae and
in coalescing neutron star binaries.
Using an energy variational principle, we show that tidal field
reduces the central density of the compact object, making it more 
stable against radial collapse. For a cold white dwarf,
the tidal field increases the maximum stable mass only slightly, 
but can actually lower the maximum central density by as much as 
$30\%$. Thus a white dwarf in a close binary may be more 
susceptible to general relativistic instability than the 
instability associated with electron capture and pycronuclear 
reaction (depending on the white dwarf composition).  
We analyse the radial stability of neutron star using
post-Newtonian approximation with an ideal degenerate neutron gas 
equation of state. The tidal stablization effect
implies that the neutron star in coalescing neutron star-neutron star
or neutron star-black hole binaries 
does not collapse prior to merger or tidal disruption.

\end{abstract}
\bigskip

\bigskip


{\bf Introduction}
\medskip

It is well-established that there exist upper
limits to the mass and central density of degenerate compact objects,
white dwarf (WD) and neutron star (NS)\cite{Zeldovich71,Shapiro83}. 
For WD, the mass asymptotes to the Chandrasekhar limit as the
central density increases, bounded by general relativistic
radial instability, electron capture 
or pycronuclear reaction. 
For NS, the existence of an absolute mass limit is truly a
general relativistic phenomena: pressure, which supports
the star against gravity, also acts as a source of gravitation. 

WD and NS frequently appear in binary systems. 
Type Ia supernova is thought to arise from accreting WD
in a close binary or the merger of a WD-WD 
binary\cite{Woosley86}. Of particular interest is the coalescing NS-NS
binaries and NS-black hole (BH) binaries, which are the most promising 
sources of gravitational waves that could be detected by 
interferometers such as 
LIGO and VIRGO\cite{Abramovici92,Cutler93,Thorne95}.
NS binary merger is also considered to be a natural engine
that drives cosmological gamma-ray bursts. 
Although the inspiral at larger orbital radius may be treated
by post-Newtonian expansion technique\cite{Blanchet95},
coupled with semi-analytic studies of the hydrodynamical 
effects\cite{Bildsten92,Kochanek92,Lai94a,Reisenegger94,Lai94,Lai95}, 
quantitative understanding of the orbital
evolution at small separation and the final merging 
requires full numerical simulation,
which still is in its
infancy\cite{Nakamura,Rasio,Davies,Centrella,Ruffert}.

The present study is motivated by the recent general relativistic 
hydrodynamical simulations 
of Wilson et al.\cite{Wilson95} which reveal evidence 
that ``general relativistic effects may cause otherwise stable neutron
stars to individually collapse prior to merging''. 
Obviously, this requires the neutron star to have a mass close to 
its maximum value to begin with (at large separation). 
However, it does raise a question as to how tidal field 
modifies the mass limit and central density limit of a 
compact object. In this paper, we analyse the radial stability 
of neutron star and white dwarf under the influence of the tidal field
of a companion. We show that the tidal effect 
generally stablizes the star.
We adopt the energy variational principle as widely used in the
analysis of stability of isolated WD and 
NS\cite{Harrison65,Zeldovich71,Shapiro83}.

\bigskip
{\bf General Consideration}
\medskip

Let $E(\rho)$ be the energy of an isolated object of 
baryon mass $M$ as a function of the central density $\rho$. 
The function $E(\rho)$ consists of internal energy and
self-gravitational energy, including their relativistic 
corrections. The equilibrium central 
density $\rho_0$ is obtained from the condition 
$(\partial E/\partial\rho)=0$; radial stability requires 
$(\partial^2E/\partial\rho^2)>0$. 

Now for the star in a binary, the dimensionless tidal distortion is of
order $\sim{\varepsilon}\equiv (M'/M)(R/r)^3$, where $R\propto
(M/\rho)^{1/3}$
is the mean stellar radius, $M'$ the companion mass, $r$ the orbital
separation. The total stellar energy can be written as 
${\cal E}(\rho)=E(\rho)+W_t(\rho)$. Note that the tidal distortion
modifies the ``intrinsic'' energy $E(\rho)$, but we will group this
correction into $W_t(\rho)$. Thus the function $W_t(\rho)$ consists of
(i) the correction $\Delta W$ to the self-gravitational potential
energy, (ii) the interaction energy $W_i$ between $M'$ and the
tide-induced quadrupole, and (iii) the kinetic energy $T_s$ of
internal fluid oscillation and rotation.
The first two contributions are of the same order $\sim
(GM^2/R){\varepsilon}^2\propto 1/r^6$, but have opposite signs,
while the third $T_s\sim MR^2{\varepsilon}^2\Omega^2\propto 1/r^9$
is a factor $(1+M'/M)(R/r)^3$ smaller, and will be neglected 
\footnote{We assume the star has
negligible viscosity to be tidally synchronized, as is the case for
compact objects\cite{Bildsten92,Kochanek92}. We also assume the star
has zero intrinsic spin --- Finite spin tends to stablize the star
against radial collapse, and can be treated separately
(e.g.,\cite{Shapiro83}). Note that there is no ambiguity in our
definition of the function ${\cal E}(\rho)$ for stability analysis: One
could add to ${\cal E}(\rho)$ a term $T=T_s+T_{orb}$ associated with the
kinetic spin and orbital energies (so that ${\cal E}$ can be
considered as the total energy of the binary system). However, 
differentiation of $T$ with respect to $\rho$ under fixed total
angular momentum $J$ and fluid circulation ${\cal C}$ gives 
$(\partial T/\partial\rho)_{J,{\cal C}}=2T_s/3\rho$, and 
the orbital motion does not affect the stellar structure.}.

To calculate $W_t$, we model the star as a polytrope
(with equation of state $P=K\rho^{1+1/n}$; $n$ is the polytropic
index). Approximating the tidaly deformed star as an ellipsoid with 
axes $a_i=R(1+\alpha_i)$ ($a_1$ is along the tidal bulge), 
the tidal distortion can be calculated in Newtonian theory.
To leading order in ${\varepsilon}$, we have\cite{Lai94b}
\begin{equation}
\alpha_1={5\over 2}q_n{M'\over M}\left({R\over r}\right)^3,~~
\alpha_2=\alpha_3=-{5\over 4}q_n{M'\over M}\left({R\over r}\right)^3,
\end{equation}
where $q_n=\kappa_n (1-n/5)$, and $\kappa_n$ is defined such 
that the moment of inertia of an isolated object is 
$I=(2/5)\kappa_n MR^2$
($\kappa_n$ specifies the mass concentration within the star). 
Note that $\alpha_1+\alpha_2+\alpha_3=0$
to leading order in ${\varepsilon}$. The two contributions to $W_t$ are
\begin{eqnarray}
\Delta W &=& {3\over 5-n}{GM^2\over R}{4\over 45}
(\alpha_1^2+\alpha_2^2+\alpha_3^2-\alpha_1\alpha_2-\alpha_2\alpha_3
\nonumber\\
&&-\alpha_1\alpha_3)={3\over 4}\kappa_nq_n{GM'^2R^5\over r^6},
\end{eqnarray}
and 
\begin{eqnarray}
W_i &=&-{GM'\over 2r^3}\left({2\over 5}\kappa_n
MR^2\right)(2\alpha_1-\alpha_2-\alpha_3)\nonumber\\
&=& -{3\over 2}\kappa_nq_n{GM'^2R^5\over r^6}.
\end{eqnarray}
Thus, up to order ${\varepsilon}^2$, the total tidal energy is 
\begin{equation}
W_t = -\lambda {GM'^2R^5\over r^6}\sim -\lambda {GM'^2\over r^6}
(M/\rho)^{5/3},
\end{equation}
with $\lambda=(3/4)\kappa_nq_n$.
Note that the negative sign in $W_t$
is crucial for the tidal stablization of radial mode. 

The equilibrium condition in the presence of tide requires
$\partial{\cal E}/\partial\rho=0$. Thus the density change
$\delta\rho=\rho-\rho_0$ (where $\rho_0$ is the equilibrium density 
of the isolated object) due to the tidal field is given by
\begin{equation}
\delta\rho={5W_t\over 3\rho}
\left({\partial^2E\over\partial\rho^2}\right)^{-1}\biggl |_{\rho_0},
\end{equation}
with the expression evaluated at $\rho_0$. Since
$(\partial^2E/\partial\rho^2)>0$ for stable configuration, we see that 
$\delta\rho<0$, i.e., {\it the tidal field reduces the central density
of a stable object}. If the star is not too close to the stability
limit, we have an estimate $\delta\rho/\rho_0\sim (-{\varepsilon}^2)$,
which is of second order in the tidal deformation ${\varepsilon}$. 
Heuristically, a tidally-distorted object is less bound gravitationally
(compared to a spherical object of the same mass), thus its volume  
expands in order to satisfy hydrostatic equilibrium. 

Now consider how the tidal field changes the maximum mass $M_m$ of the
object. Let $\rho_{m0}$ and $M_{m0}$ be the zero-tide values of the
maximum density and maximum mass, at which $\partial E/\partial\rho
=\partial^2E/\partial\rho^2=0$ is satisfied. Taking the difference
between $(\partial{\cal E}/\partial\rho)|_{\rho_m,M_m}=0$ and
$(\partial E/\partial\rho)|_{\rho_{m0},M_{m0}}=0$, we obtain
\begin{equation}
\delta M_m=M_m-M_{m0}={5W_t\over 3\rho}
\left({\partial^2E\over\partial\rho\partial M}
\right)^{-1}\!\biggl |_{\rho_{m0},M_{m0}}.
\end{equation}
Since we can show\footnote{Consider a sequence of stellar models
parametrized by the central density. Differentiating 
$\partial E/\partial\rho=0$ we obtain:
$(\partial^2E/\partial\rho^2)+(\partial^2E/\partial\rho\partial M)
(dM/d\rho)=0$. The stable branch of the sequence satisfies
$(\partial^2E/\partial\rho^2)>0$ and $dM/d\rho>0$, while the
unstable branch satisfies $(\partial^2E/\partial\rho^2)<0$ 
and $dM/d\rho <0$. In both cases we have
$(\partial^2E/\partial\rho\partial M)<0$.}
$(\partial^2E/\partial\rho\partial M)<0$, we find
$\delta M_m>0$, i.e., {\it the tidal field increases the maximum mass
for stability}. An order of magnitude estimate gives
$\delta M_m/M_{m0}\sim \lambda\,{\varepsilon}^2$. 

Note that two implicit assumptions have been made in our analysis:
(i) The star ``relaxes'' to its equilibrium shape even when the 
density is out of equilibrium. This separation of the radial 
motion and the ``shape adjustment'' makes our analysis of radial 
instability more transparent. It is valid because near the 
stability limit, the radial oscillation has almost zero frequency. 
(ii) The displacement of a fluid element inside the star is a
linear function of the fluid position. This is exact
only for the $n=3$ polytrope and in the incompressible limit, 
but otherwise corresponds to an approximate trial wavefunction in the
variational principle. 


\bigskip
{\bf White Dwarf}
\medskip

We now consider the stability of white dwarfs. 
Near the maximum mass, the stellar density profile 
resembles that of a $n=3$ polytrope. The ``intrinsic'' energy 
can be written as\cite{Zeldovich71,Shapiro83}
\begin{equation} 
E=E_{int}+W+\Delta E_{int}+\Delta E_{GR},
\end{equation}
where $E_{int}\sim M\rho^{1/3}$ is the
internal energy of the ultra-relativistic electrons, 
$\Delta E_{int}\sim M\rho^{-1/3}$ is the correction due to
finite electron mass, $W\sim -M^2/R\sim -M^{5/3}\rho^{1/3}$
is the potential energy of self-gravity, and 
$\Delta E_{GR}\sim -(M^2/R)(M/R)\sim -M^{7/3}\rho^{2/3}$
is the general relativistic (post-Newtonian) correction. 
For isolated cold WD, this gives for the maximum central density and
maximum mass $\rho_{m0}=2.737\times 10^{10}$ g\,cm$^{-3}$,
$M_{m0}=1.4156M_\odot$, the correcponding minimum radius
$R_{m0}=1110$ km (We assume that the number of electron per nucleon
is $Y_e=0.5$).

We parametrize the strength of the tidal field by the 
dimensionless ratio $\beta\equiv (M'/M_{m0})(R_{m0}/r)^3$.
To avoid tidal disruption we require 
$\beta\mathrel{\raise.3ex\hbox{$<$}\mkern-14mu\lower0.6ex\hbox{$\sim$}}
0.1$. With $\lambda\simeq 0.01$ (for $n=3$ polytrope) we then have
\begin{equation}
W_t\simeq -0.01\,\beta^2\,{GM_{m0}^2\over R_{m0}}
\left({M\over M_{m0}}\right)^{5/3}\!
\left({\rho\over\rho_{m0}}\right)^{-5/3}.
\end{equation}
Expressing mass in the units of $M_\odot$, density
in $10^{10}$ g\,cm$^{-3}$, energy in $10^{51}$ erg,
the total energy can be written as
\begin{eqnarray}
{\cal E} &=& 3.7129M\rho^{1/3}-2.8895 M^{5/3}\rho^{1/3}
+0.0457M\rho^{-1/3}\nonumber\\
&&-0.0105M^{7/3}\rho^{2/3}-0.14\,\beta^2(M/\rho)^{5/3}.
\end{eqnarray}

Solving $(\partial{\cal E}/\partial\rho)=(\partial^2{\cal E}
/\partial\rho^2)=0$
yields the maximum density and maximum mass as a function of 
$\beta$. The result is shown in Figure 1. 
The increase in the maximum mass is rather small
($\mathrel{\raise.3ex\hbox{$<$}\mkern-14mu\lower0.6ex\hbox{$\sim$}}
0.1\%$). However, {\it the maximum density decrease can be
rather substantial, reaching as much as $\sim 30\%$}.
For small $\beta$, we have $\delta M_m/M_{m0}\simeq 0.05\beta^2$, 
and $\delta\rho_m/\rho_{m0}\simeq -17\,\beta^2$.
Figure 1 also shows the central density of three constant-mass
sequences with mass slightly below $M_{m0}$. We see that as
the binary separation decreases ($\beta$ increases), the central
densities also decrease, and always remain smaller than the maximum 
density $\rho_m(\beta)$ allowed for radial stability.

The fact that the maximum density $\rho_m(\beta)$ for stability 
decreases with decreasing $r$ may have some interesting astrophysical
consequences. The maximum central density of a normal Carbon-Oxygen
white dwarfs is set by the threshold of electron capture on 
$^{16}$O, at $\rho_{cap}=1.9\times 10^{10}$ g\,cm$^{-3}$
(the pycronuclear reaction between $^{12}$C may set in at smaller
density, of order $10^{10}$ g\,cm$^{-3}$; but this increases
as the $^{12}$C abundance decreases\cite{Sahrling94}).
From Fig.~1 we see that $\rho_m(\beta)$ can drop below $\rho_{cap}$,
therefore GR effect becomes more important than neutronization.
This implies that an accreting white dwarf in close binary 
can be more susceptible to collapsing to NS by general 
relativistic radial instability. Of course, 
the real situation may be complicated by the finite temperature 
due to the high accretion rate. 

\bigskip
{\bf Neutron Star}
\medskip

The equation of state of nuclear matter is uncertain. For
the purpose of illustration, we adopt EOS to be that of
an ideal degenerate neutron gas and analyse the radial stability 
in the post-Newtonian approximation. 
Assuming the stellar density profile to be that of the $n=3/2$ 
polytrope, the ``intrinsic'' energy can  still be written as
in Eq.~(7), except that $E_{int}\sim M\rho^{2/3}$ is the internal
energy of nonrelativistic neutron, $\Delta E_{int}\sim -M\rho^{4/3}$
is the correction due to special relativistic effect. 
The maximum baryon mass of an isolated NS bassed on this model is 
$M_{m0}=1.1108M_\odot$, the maximum central density 
$\rho_{m0}=7.415\times 10^{15}$ g\,cm$^{-3}$, and the corresponding 
minimum radius $R_{m0}=10.48$ km. 
We again parametrize the tidal strength 
by the ratio $\beta\equiv (M'/M_{m0})(R_{m0}/r)^3$.
The tidal energy can then be written in the form
of Eq.~(8) except that we use $\lambda\simeq 0.1$ appropriate 
for $n=3/2$ polytrope (The uncertainty can be readily absorbed into
the definition of $\beta$). The GR correction to the tidal potential
energy is of order $(M'/r)$ smaller, and is neglected. 
Expressing mass in the units of $M_\odot$, density
in $10^{15}$ g\,cm$^{-3}$, and energy $10^{53}$ erg,
the total energy function can be written as
\begin{eqnarray}
{\cal E} &=& 0.85168 M\rho^{2/3}-1.5968 M^{5/3}\rho^{1/3}
-0.02887M\rho^{4/3}\nonumber\\
&&-0.16774 M^{7/3}\rho^{2/3}-7.356\,\beta^2(M/\rho)^{5/3}.
\end{eqnarray}

Figure 2 shows the maximum density $\rho_m$ and maximum mass 
$M_m$ as a function of $\beta$. We see that, as in the white dwarf 
case, $\rho_m$ decreases and $M_m$ increases with increasing $\beta$. 
For small $\beta$ 
($\mathrel{\raise.3ex\hbox{$<$}\mkern-14mu\lower0.6ex\hbox{$\sim$}}
0.15$), we have 
\begin{eqnarray}
{\delta M_m\over M_{m0}} &\simeq& -{15W_t\over 4(2E_{in}+W)}
=0.33\,\beta^2,\\
{\delta \rho_m\over \rho_{m0}} &\simeq& -{5W_t(28E_{in}+15W)
\over 2W(2E_{in}+W)}=-2.7\,\beta^2.
\end{eqnarray}
Figure 2 also depicts the densities of
constant-mass sequences with $M$ slightly below $M_{m0}$. 
Again, we see that these densities $\rho(\beta)$ always 
remain below $\rho_m(\beta)$, i.e., these stars are stable
against radial perturbation until the tidal limit $\beta\simeq 0.23$
is reached. 

Another way to look at the tidal effect on the radial stability 
is to construct sequence of stellar models with varying central
densities at a given fixed $\beta$. Figure 3 shows the $M-\rho$ 
curves of several such sequences. Only models in the $dM/d\rho>0$ 
branch are stable, and the maximum density $\rho_m$ and maximum mass 
$M_{m}$ are given by the values at the turning point $dM/d\rho=0$.
One can readily see that the effects of the tidal field is to raise
$M_m$ and lower $\rho_m$. For 
$\beta\mathrel{\raise.3ex\hbox{$>$}\mkern-14mu\lower0.6ex\hbox{$\sim$}}
0.23$, no stable configuration
exists, and this corresponds to the tidal disruption limit.

\bigskip
{\bf Discussions}
\medskip

The above analysis demonstrates that tidal effect
tends to stabilize neutron star and white dwarf against
radial collapse, at least within the framework of post-Newtonian
theory. Our method is accurate for white dwarfs, but only 
approximate for neutron stars. However,
we think it is unlikely that higher order GR corrections or 
the use of more sophisticated nuclear equation of state\cite{Wilson95} 
will change our qualitative results for neutron stars, 
although the precise numbers can certainly change. 

In the cases of NS-NS binaries, one might still consider the
possibility of neutron star collapse prior to merging when there is
stable mass transfer from its lower mass companion. However, by the
time mass transfer starts, the binary must already have
encountered the orbital dynamical instability as a result of strong
tidal interaction\cite{Lai94a}. This tidal instability,
enhanced by general relativistic effects\cite{Kidder93}
leads to rapid coalescence of the binary within a few
orbits\cite{Lai95,Lai96}.

\bigskip

I thank Kip Thorne for comments on an earlier version of the 
paper and helpful suggestions. 
This research is supported by the Richard C. Tolman Fellowship 
in theoretical astrophysics at Caltech, and NSF Grant AST-9417371
and NASA Grant NAG 5-2756.


\bigskip
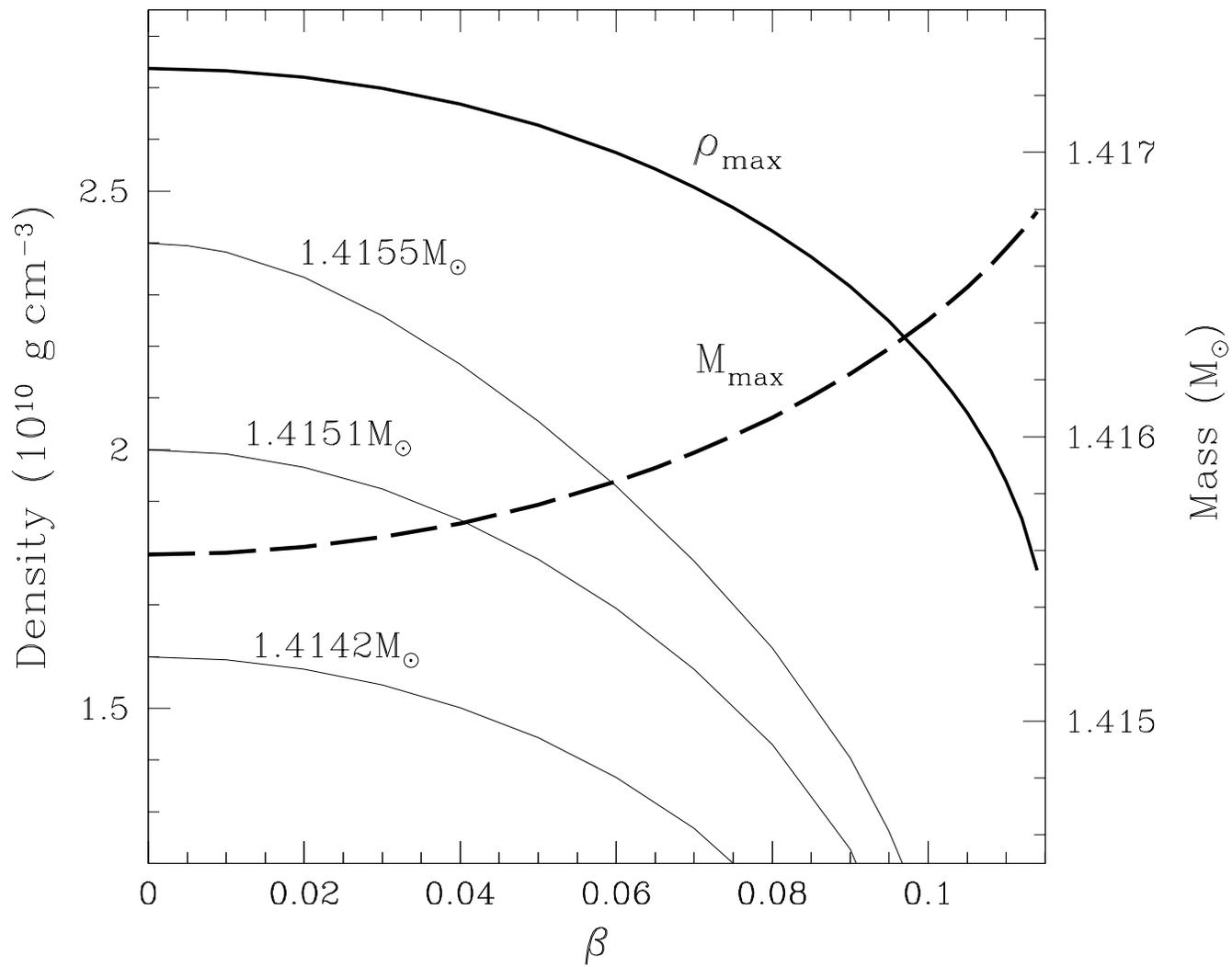
\begin{figure}
\caption
{The maximum central density (heavy solid line)
and maximum mass (dashed line) of white dwarf as a function
of the tidal ratio $\beta\equiv (M'/M_{m0})(R_{m0}/r)^3$. 
The solid lines show the central densities of three sequences
with mass slightly smaller than maximum mass of a isolated
WD $M_{m0}=1.4156M_\odot$. 
}
\end{figure}

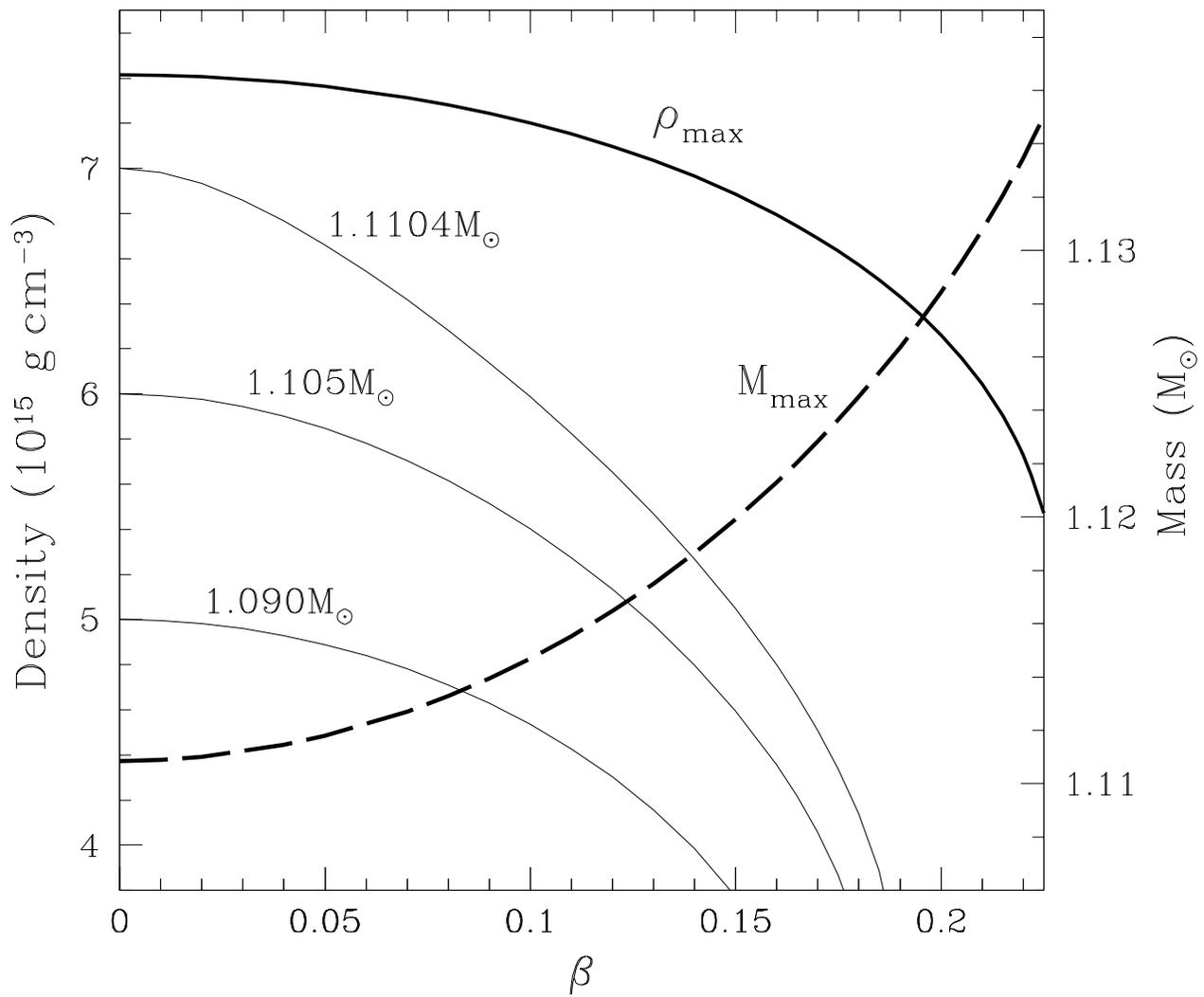
\begin{figure}
\caption
{The maximum central density (heavy solid line)
and maximum mass (dashed line) of neutron star as a function
of the tidal ratio $\beta\equiv (M'/M_{m0})(R_{m0}/r)^3$. 
The solid lines show the central densities of three sequences
with mass slightly smaller than maximum mass of an isolated
NS $M_{m0}=1.1108M_\odot$. 
}
\end{figure}

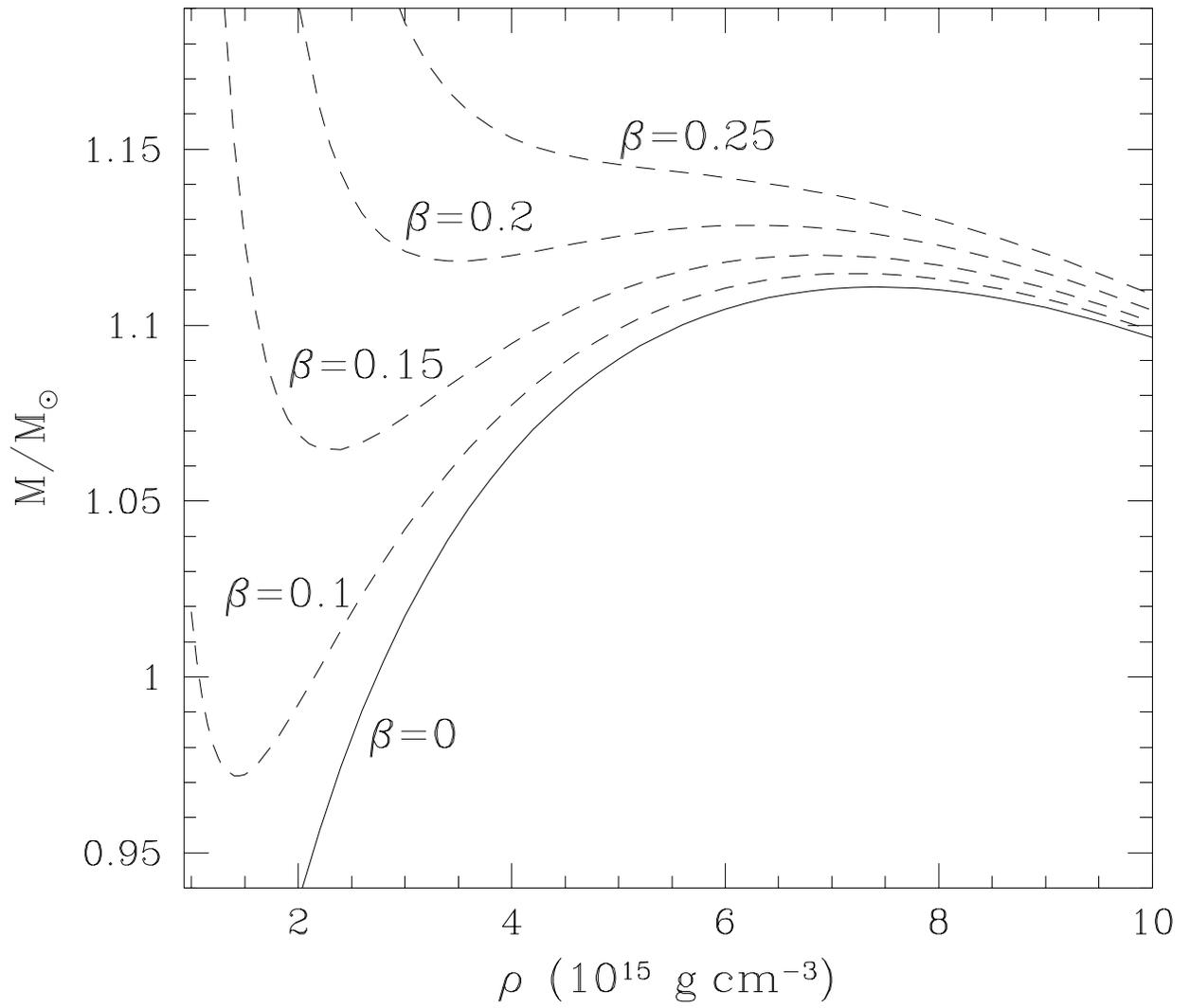
\begin{figure}
\caption
{The mass-central density curves of sequences of NS models with
constant values of $\beta$. ($\beta=0$ corresponds to isolated NS). 
}
\end{figure}

\end{document}